\documentclass[aps,prl,twocolumn, superscriptaddress,showpacs]{revtex4-1}

\usepackage[T1]{fontenc}
\usepackage[latin1]{inputenc}
\usepackage[dvips]{graphicx,color}
\usepackage{rotating}
\usepackage{bm}        
\usepackage{amssymb}   
\usepackage{amsmath} 
\def\jcp#1#2#3{J.~Chem.~Phys.~{\bf #1},\ #2\ (#3)}

\def\pra#1#2#3{Phys.~Rev.~A~{\bf #1},\ #2\ (#3)}
\def\prl#1#2#3{Phys.~Rev.~Lett.~{\bf #1},\ #2\ (#3)}

\def\k1{k_1}
\def\k2{k_2}
\def\q1{q_1}
\def\q2{q_2}

\def\({\left (}
\def\){\right )}
\def\[{\left [}
\def\]{\right ]}

\newcommand{\beq}{\begin{equation}}
\newcommand{\eeq}{\end{equation}}

\begin{document}
\date{\today}
\title{Sympathetic cooling of polyatomic molecules with $S$-state atoms in a magnetic trap}

\author{T. V. Tscherbul}
\affiliation{Harvard-MIT Center for Ultracold Atoms, Cambridge, Massachusetts 02138}
\affiliation{Institute for Theoretical Atomic, Molecular and Optical Physics,
Harvard-Smithsonian Center for Astrophysics, Cambridge, Massachusetts 02138}
\author{H-G. Yu}
\affiliation{Chemistry Department, Brookhaven National Laboratory, Upton, New York 11973}
\author{A. Dalgarno}
\affiliation{Harvard-MIT Center for Ultracold Atoms, Cambridge, Massachusetts 02138}
\affiliation{Institute for Theoretical Atomic, Molecular and Optical Physics, 
Harvard-Smithsonian Center for Astrophysics, Cambridge, Massachusetts 02138}

\begin{abstract}
We present a rigorous theoretical study of low-temperature collisions of polyatomic molecular radicals with $^1S_0$ atoms in the presence of an external magnetic field. Accurate quantum scattering calculations based on {\it ab initio} and scaled interaction potentials show that collision-induced spin relaxation of the prototypical organic molecule CH$_2(\tilde{X}^3B_1)$ (methylene)  and 9 other triatomic radicals in cold $^3$He gas occurs at a slow rate, demonstrating that cryogenic buffer-gas cooling and magnetic trapping of these molecules is feasible with current technology. Our calculations further suggest that it may be possible to create ultracold gases of polyatomic molecules by sympathetic cooling with alkaline-earth atoms in a magnetic trap.

\end{abstract}

\maketitle

\clearpage
\newpage

The study of molecular structure and interactions at low temperatures is a promising route toward exploring novel phenomena in physics and chemistry, with applications ranging from quantum simulation of strongly correlated many-body systems \cite{NJP} to quantum information processing \cite{Andre} and metrology \cite{Metrology}. Recent experimental observation of chemical reactions in ultracold molecular gases \cite{KRb} and atom-molecule mixtures \cite{Jose} has opened a new chapter in chemical physics by demonstrating the possibility of controlling bimolecular chemical reactions with external electromagnetic fields \cite{Roman05}. Some of the most important experimental techniques for the creation of cold molecules include photoassociation of ultracold atoms \cite{NJP,KRb}, Stark deceleration of molecular beams \cite{StarkDeceleration}, and cryogenic buffer gas cooling \cite{BufferGasCooling,HeNH}. 


With few exceptions \cite{Ammonia,H2CO}, previous experimental and theoretical work has been limited to diatomic molecules \cite{NJP}. Cooling and trapping of polyatomic molecules would greatly enrich the scope of molecular physics and allow the study of complex chemical reactions \cite{Sims}, molecular decoherence \cite{C60} and parity violation \cite{ParityViolation} in the well-controlled environment of an electromagnetic trap. A potentially viable technique for creating ultracold polyatomic molecules could be based on cryogenic buffer gas cooling \cite{BufferGasCooling} or Stark deceleration \cite{H2CO} followed by magnetic trapping and subsequent sympathetic cooling with ultracold atoms \cite{MgNH,Barker,N-NH}. Low-temperature collisional properties of polyatomic molecules play a key role in this scheme: only the molecules in low-field-seeking Zeeman states can be confined in a magnetic trap, and such molecules are vulnerable to spin relaxation induced by collisions with buffer gas atoms, causing trap loss. For efficient sympathetic cooling, the number of elastic collisions per inelastic collision $\gamma$ must exceed 100 \cite{NJP,BufferGasCooling}.

In this Letter, we use rigorous quantum scattering calculations to demonstrate the attainability of sympathetic cooling of 10 polyatomic molecular radicals with cold $^3$He atoms in a magnetic trap. To our knowledge, this is the first rigorous theoretical analysis of low-temperature collisions of polyatomic molecular radicals that incorporates in a rigorous manner both intra- and intermolecular interactions and the effects of an external magnetic field on collision dynamics. Our calculations provide compelling evidence that methylene (CH$_2$)---the prototypical organic molecule---can be cooled and magnetically trapped by collisions with cryogenic He buffer gas, thereby opening up novel opportunities for research in astrophysics, combustion, organic chemistry, molecular interferometry, and precision spectroscopy. In addition, our study provides insight into the mechanisms of spin relaxation of polyatomic molecules in collisions with $S$-state atoms, demonstrating that polyatomic molecules with one unpaired electron undergo inelastic spin relaxation much slower than molecules with two unpaired electrons. Our calculated low-temperature collision rates for polyatomic molecules are similar to those measured previously for the diatomic molecules CaH($^2\Sigma$) and NH($^3\Sigma$), which indicates that advanced collisional cooling techniques that are currently under development for diatomic molecules \cite{MgNH,Barker,N-NH} can be extended in a straightforward manner to polyatomic molecules, thereby demonstrating a pathway to ultracold ensembles of chemically diverse molecules that cannot be created by any other cooling method in existence today.

We begin by specifying the Hamiltonian for an open-shell polyatomic molecule in the ground electronic and vibrational states colliding with a $^1S_0$ atom in the presence of an external magnetic field ($\hbar=1$)
\begin{equation}\label{H}
\hat{H} = -\frac{1}{2\mu R}\frac{\partial^2}{\partial R^2}R + \frac{\hat{\ell}^2}{2\mu R^2} + V(\bm{R},\hat{\Omega}) + \hat{H}_\text{mol},
\end{equation}
where $\mu$ is the reduced mass of the collision complex, $\bm{R}$ is the atom-molecule separation vector of length $R$, $\hat{\ell}^2$ is the orbital angular momentum for the collision and $\hat{H}_\text{mol}$ describes the internal structure of the molecule 
\begin{equation}\label{Hmol}
\hat{H}_\text{mol} = A\hat{N}_x^2 +  C\hat{N}_y^2 + B\hat{N}_z^2 + \hat{H}_\text{cd}  + \hat{H}_\text{sr} + \hat{H}_\text{ss} + \hat{H}_\text{ext} 
\end{equation}
where $A$, $B$, and $C$ are the rotational constants of an asymmetric top, $\hat{N}_\alpha$ ($\alpha = x,y,z$) yield the components of rotational angular momentum $\hat{N}$ in the molecule-fixed (MF) coordinate frame. We identify the MF axes with the principal axes of inertia of the molecule \cite{Hutson}. The spin-rotation interaction is given by
\cite{Bunker}
\begin{multline}\label{Hsr}
\hat{H}_\text{sr} = \bar{\gamma}\hat{N}\cdot\hat{S} + \sum_{q=-2}^2 \biggl{[} \frac{1}{2} (\gamma_x-\gamma_y)  [\mathcal{D}^2_{q2}(\hat{\Omega}) + \mathcal{D}^2_{q,-2}(\hat{\Omega})] \\ +    \frac{1}{\sqrt{6}}(2\gamma_z-\gamma_x-\gamma_y) \mathcal{D}^2_{q0}(\hat{\Omega}) \biggr{]} [\hat{N}\otimes\hat{S}]^{(2)}_q
\end{multline}
where $\hat{S}$ is the electron spin, $\gamma_\alpha$ are the spin-rotation constants, $\bar{\gamma}=\frac{1}{3}(\gamma_x+\gamma_y+\gamma_z)$, $\mathcal{D}^2_{qq'}(\hat{\Omega})$ are the Wigner $D$-functions, and $\hat{\Omega}$ are the Euler angles which specify the orientation of MF axes in the space-fixed (SF) coordinate frame. The spin-spin interaction in $S=1$ molecules may be written as \cite{Bunker,TBP}
\begin{multline}\label{Hss}
\hat{H}_\text{ss} = \sum_{q=-2}^2 \biggl{[} \frac{1}{2}(E+D) [\mathcal{D}^2_{q2}(\hat{\Omega}) + \mathcal{D}^2_{q,-2}(\hat{\Omega})] \\+  \frac{1}{\sqrt{6}}(3E-D) \mathcal{D}^2_{q0}(\hat{\Omega}) \biggr{]} [\hat{S}\otimes\hat{S}]^{(2)}_q
\end{multline}
where $D$ and $E$ are the zero-field-splitting parameters \cite{Bunker,note}. The interaction of the molecule with an external magnetic field is described by $\hat{H}_\text{ext} = 2\mu_0 B \hat{S}_Z$, where $B$ is the magnetic field strength, $\mu_0$ is the Bohr magneton and $\hat{S}_Z$ yields the SF projection of $\hat{S}$. The term $\hat{H}_\text{cd}$ accounts for the effects of centrifugal distortion \cite{Bunker}.

To solve the quantum scattering problem for the Hamiltonian (\ref{H}), we expand the wave function of the collision complex in the fully uncoupled SF basis $|NM_NK_N\rangle |SM_S\rangle |\ell m_\ell\rangle$, where $|NM_NK_N\rangle$ are the symmetric top eigenfunctions, $M_N$, $m_\ell$, and $M_S$ are the SF projections of $\hat{N}$, $\hat{\ell}$, and $\hat{S}$, and $K_N$ is the MF projection of $\hat{N}$. This expansion leads to a system of coupled-channel (CC) equations parametrized by the matrix elements of the Hamiltonian (\ref{H}), which can be evaluated analytically \cite{TBP} using the known spectroscopic constants of polyatomic radicals \cite{Bunker,TBP}. The matrix elements of the atom-molecule interaction potential can be evaluated by expanding the potential in angular functions \cite{Hutson,TBP}
\begin{multline}\label{V}
V(\bm{R},\hat{\Omega}) = \sum_{\lambda,\mu\ge 0} V_{\lambda\mu} (R) \frac{1}{1+\delta_{\mu 0}} \\ \times \sum_{m_\lambda} [\mathcal{D}^\lambda_{m_\lambda\mu}(\hat{\Omega}) + (-)^\mu \mathcal{D}^\lambda_{m_\lambda,-\mu}(\hat{\Omega})] Y_{\lambda m_\lambda}(\hat{R})
\end{multline}
where $\hat{R}=\bm{R}/R$. 

\begin{figure}[t]
	\centering
	\includegraphics[width=0.43\textwidth, trim = 0 0 0 0]{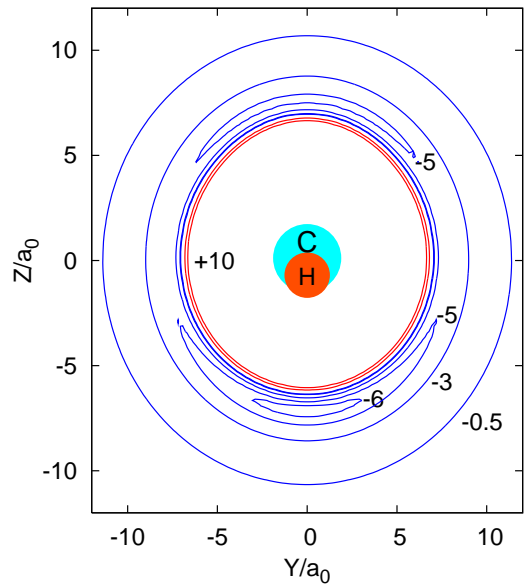}
	\renewcommand{\figurename}{Fig.}
	\caption{A contour plot of the {\it ab initio} He-CH$_2$ PES in Cartesian coordinates $X_\text{MF} = R\sin\theta\cos\phi$ and $Z_\text{MF} = R\cos\theta$ for $\phi=90^\circ$. Energies are in cm$^{-1}$ (1 cm$^{-1} = 1.439$ K).}\label{fig:pes}
\end{figure}


At this point, we specialize our analysis to the He~+~CH$_2$ collision system. CH$_2$ (methylene) is the simplest polyatomic molecular radical with a triplet ground state of $\tilde{X}^3B_1$ symmetry, which serves as a prototype for studying the mechanisms of inelastic collisions and chemical reactions of polyatomic molecules \cite{Hall}. Methylene has recently been detected in interstellar space \cite{AstroCH2}, and in Titan's atmosphere, where it is formed by photodissociation of methane \cite{Titan}. In addition, CH$_2$ is the simplest known carbene, an important class of reaction intermediates in organic chemistry \cite{CarbenesBook}.

 The ground rotational state of CH$_2$ is split by an external magnetic field into three Zeeman sublevels with $M_S = 0$ and $\pm 1$. Like homonuclear molecules with non-zero nuclear spins, methylene can exist in two spin modifications ($ortho$ and $para$), only one of which ($o$-CH$_2$) is amenable to magnetic trapping \cite{TBP}. In this work, we consider collisions of $o$-CH$_2$ molecules in their maximally spin-stretched Zeeman sublevel of the ground rotational state $|M_S=1\rangle$. Collisions with He atoms induce spin relaxation transitions to the Zeeman states $|M_S=0\rangle$ and $|M_S=-1\rangle$, which are the main focus of this work. To quantify the dynamics of spin relaxation, we solve the CC equations numerically on a grid of $R$ from 2 to 40$a_0$ with a grid step of 0.04 $a_0$,  The CC basis set included five rotational states of CH$_2$, and six partial wave states ($\ell = 0-5$), leading to 1123 CC equations for $M=0$. The calculated cross sections were converged to $<$10\%.

To evaluate the potential energy surface (PES) for the He-CH$_2$ collision complex, we use a high-level {\em ab initio} approach based on the coupled-cluster method with single, double, and non-iterative triple excitations [CCSD(T)] \cite{WAT93:8718} and a large basis set of cc-pVTZ quality \cite{WOO93:1358}. A total of 985 {\it ab initio} points were calculated and fitted by an analytic function consisting of six sets of long-range pair potentials.  The root-mean-square error of the fit was 0.8~cm$^{-1}$. A contour plot of the calculated PES in shown in Fig.~\ref{fig:pes}. The global minimum is located at $R=7.14\,a_0$ and $\theta=\pi$ and has a well depth of 6.3\,cm$^{-1}$.  The energy difference between the linear and T-shaped configurations amounts to $\sim$1.5 cm$^{-1}$ for $\phi = 90^\circ$, demonstrating that the He-CH$_2$ interaction is weakly anisotropic.

\begin{figure}[t]
	\centering
	\includegraphics[width=0.43\textwidth, trim = 0 0 0 0]{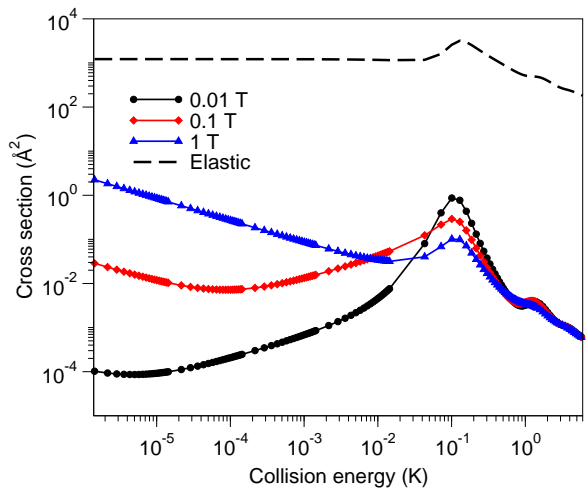}
	\renewcommand{\figurename}{Fig.}
	\caption{($a$) Cross sections for elastic scattering (dashed line) and spin relaxation in He + CH$_2$ collisions plotted vs collision energy for different magnetic field strengths: 0.01~T (circles), 0.1~T (diamonds), and 1~T (triangles). The $s$-wave scattering lengths for $B=0.1$~T are $18.6$ ($f_s=1$) and 13.9 $a_0$ ($f_s=10$).}\label{fig:xs}
\end{figure}

Figure~\ref{fig:xs} shows the cross sections for elastic scattering and spin relaxation in He~+~CH$_2$ collisions as functions of collision energy. 
A pronounced maximum near $E_C \sim 0.1$~K occurs due to a shape resonance in the incident collision channel. At ultralow collision energies, the cross sections for elastic energy transfer approach a constant value and those for spin relaxation vary as $E_C^{-1/2}$, according to the Wigner threshold law for $s$-wave scattering.  The sensitivity of the cross sections to magnetic field is the strongest at ultralow collision energies, and near the shape resonance at $E_C\sim 0.1$~K.

\begin{figure}[t]
	\centering
	\includegraphics[width=0.37\textwidth, trim = 0 0 0 0]{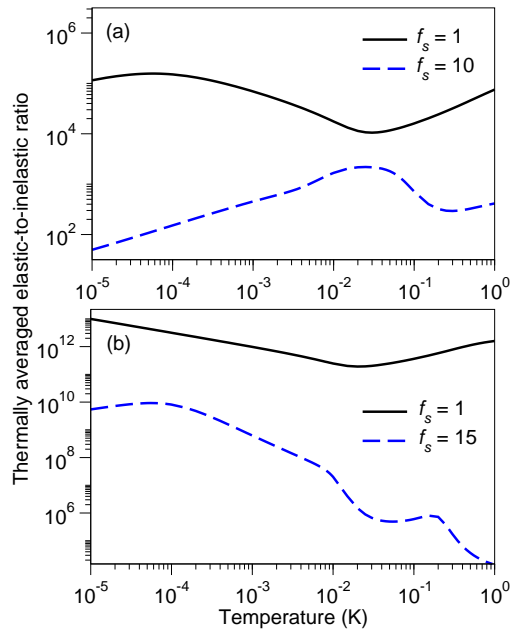}
	\renewcommand{\figurename}{Fig.}
	\caption{Ratios of elastic to inelastic collision rates $\gamma(T)$ for (a) He + CH$_2$ and (b) He + NH$_2$ as functions of temperature. Full lines: results for the unmodified He-CH$_2$ interaction potential ($f_s=1$), dashed lines: results for the scaled potential \cite{SI}.  The magnetic field is 0.1~T.}\label{fig:gammas}
\end{figure}

Figure~\ref{fig:gammas}(a) shows the calculated ratio of the rate constants for elastic scattering and spin relaxation in He~+~CH$_2$ collisions at a magnetic field of 0.1 T. The ratio remains high ($\gamma>10^4$) over a wide range of temperatures from 1 K down to 10$^{-5}$ K. In order to examine the sensitivity of these results to the anisotropic part of the interaction potential, we repeated scattering calculations with a modified He-CH$_2$ interaction potential obtained by scaling the $V_{\lambda\mu}(R)$ terms with $\lambda>0$ (Eq. \ref{V}) by a constant factor $f_s=10$. The results plotted in Fig.~\ref{fig:gammas}(a) demonstrate that even for this significantly more anisotropic interaction potential, $\gamma(T)$ hardly ever falls below 100, demonstrating that CH$_2$ molecules trapped in the presence of cold $^3$He gas will be extremely stable against collisional spin relaxation.

In order to explore the possibility of sympathetic cooling and magnetic trapping of polyatomic molecules other than CH$_2$, we have extended our analysis to include the NH$_2$ molecule in its ground rovibronic state of $^2A$ symmetry. Figure~\ref{fig:gammas}(b) shows the elastic-to-inelastic ratio for He~+~NH$_2$ collisions computed using the He-CH$_2$ interaction potential and the accurate spectroscopic constants of NH$_2$ \cite{TBP}. The interactions of $\Sigma$-state molecules with He atoms are generally weakly anisotropic \cite{NJP,Roman03}, so we believe that the replacement of the unknown He-NH$_2$ interaction PES by the He-CH$_2$ PES computed in this work is a reasonable approximation.  The elastic-to-inelastic ratio for He~+~NH$_2$ is remarkably high, exceeding that for He~+~CH$_2$ by nearly 7 orders of magnitude. Calculations with the modified PES yield $\gamma \sim10^4$, still perfectly acceptable in the context of buffer gas cooling and magnetic trapping experiments \cite{BufferGasCooling}.

Table I lists the values of $\gamma$ for 18 molecular radicals obtained from quantum scattering calculations using the accurate molecular constants \cite{TBP} and both the original and modified He-CH$_2$ PESs. Because the magnitude of $\gamma$ can be sensitive to the anisotropic part of the PES \cite{HeNH}, the $f_s=1$ results are best thought of as order-of-magnitude estimates, whereas the $f_s\ge 10$ results may be regarded as conservative lower bounds \cite{SI}. A few important qualitative conclusions can be drawn based on the results listed in Table~I. All $S=1/2$ molecules except HCO have extremely large elastic-to-inelastic ratios, which makes them amenable to sympathetic cooling and magnetic trapping experiments using cold $^3$He gas. This conclusion is independent of whether unmodified ($f_s=1$) or strongly anisotropic ($f_s\ge 10$) interaction potentials are used in scattering calculations, even though the magnitude of $\gamma$ is sensitive to $f_s$. Apart from CH$_2$, only one triplet radical (CHF) satisfies the criterion $\gamma>100$ for $f_s=10$. To explain this result, we note that spin relaxation in $S=1$ molecules is driven by the spin-spin interaction (\ref{Hss}), which couples the ground initial Zeeman state $|N=0, M_S=1\rangle$ to a manifold of rotationally excited states $|N=2,M_S\rangle$. These couplings lead to direct spin-flipping transitions, whose cross sections scale quadratically with $D$ \cite{HeNH}. In the absence of the spin-spin interaction, spin relaxation occurs via a much less efficient indirect mechanism \cite{Roman03}.

\begin{table}[t!]
\caption{Ratios of the rate constants for elastic energy transfer and spin relaxation calculated for 18 polyatomic molecular radicals with $^3$He using unmodified $(f_s=1)$ and strongly anisotropic ($f_s=10$ unless noted otherwise \cite{SI}) interaction potentials for He-CH$_2$. $T=0.5$~K and $B=0.1$~T.} 
\centering
\vspace{0.3cm}

\begin{tabular}{cccccc}
\hline
\hline
Molecule &  \multicolumn{2}{c}{$\gamma$} &  Molecule & \multicolumn{2}{c}{$\gamma$}  \\
       ($S=1$)                 &  $f_s = 1$   &   $f_s=10$        & ($S=\frac{1}{2}$)  &  $f_s = 1$    &   $f_s = 10$  \\
\hline
CH$_2$  &      $4.5\times 10^{4}$   & 326 & NH$_2$  & $1.1\times 10^{12}$  & $3.6\times 10^{4}$\\
CHF       &      $1.7\times 10^{3}$   & 186 & PH$_2$  & $4.2\times 10^{10}$  & $6.3\times 10^{3}$ \\
CHCl      &      $418$                     & 11 & AsH$_2$ & $9.5\times 10^{9}$  & $140$ \\
CHBr      &      $156$                     & 3 & HO$_2$   & $1.8\times 10^{9}$  & $5.6\times 10^{3}$ \\
CHI        &      $64$                       & 10 & HCO       & $1.9\times 10^{11}$  & $25$ \\
CF$_2$  &      $80$                       & 11 & NF$_2$  & $4.6\times 10^{7}$   & 894 \\
CCl$_2$  &      $11$                      & 9 & NO$_2$  & $2.1\times 10^{8}$   & 541\\
CFCl      &      $10$                       & 8 & ClO$_2$  & $4.9\times 10^{4}$  & 260 \\
SiH$_2$  &      $2.5\times 10^{4}$   & 34 & CH$_3$  & $9.8\times 10^{12}$  & $2.8\times 10^{7}$ \\
\hline
\hline
\end{tabular}\begin{flushleft}
\end{flushleft}
\end{table}

In summary, we have presented a rigorous theoretical analysis of low-temperature collisions of polyatomic molecular radicals with closed-shell atoms in the presence of an external magnetic field using an accurate {\it ab initio} interaction potential. The calculations demonstrate that spin relaxation in $^3$He~+~CH$_2$ collisions occurs at a remarkably slow rate of $1.2\times 10^{-14}$ cm$^3$/s at $T= 0.5$ K, thereby making CH$_2$($\tilde{X}^3B_1$) an ideal candidate for sympathetic cooling experiments using cold $^3$He gas. We have also presented model calculations of spin relaxation rates for other polyatomic molecules, demonstrating that two $S=1$ molecules and eight  $S=1/2$ molecules listed in Table I should be amenable to cryogenic buffer-gas cooling and magnetic trapping with long lifetimes.

This work demonstrates the feasibility of sympathetic cooling and magnetic trapping of a large class of molecular radicals of importance in organic chemistry, astrophysics, precision spectroscopy, and molecular interferometry, which may open up novel avenues of research. Inelastic collisions, chemical reactions and three-body recombination of polyatomic molecules can now be studied in cold collision experiments \cite{BufferGasCooling,Roman05}. Accurate measurements of radiative lifetimes and reaction rates of co-trapped molecular radicals would greatly facilitate astrochemical modeling of dense cold interstellar clouds \cite{Sims,AstroCH2}. Magnetic trapping of polyatomic molecules may provide  novel routes to probing quantum decoherence \cite{C60} and controlling reaction mechanisms with external fields \cite{Roman05}. Finally, we note that the calculated elastic-to-inelastic ratios for molecular species listed in Table I are similar in magnitude to those already measured for CaH($^2\Sigma$) \cite{Roman03} and NH($^3\Sigma$) \cite{HeNH}, which indicates that it may be possible to create ultracold ensembles of polyatomic molecules via sympathetic cooling with laser-cooled alkaline earth or spin-polarized N atoms \cite{MgNH,Barker,N-NH}.

We thank G.E. Hall and J.M. Doyle for discussions. This work was supported by the DOE Office of Basic Energy Science and NSF grants to the Harvard-MIT CUA and ITAMP at Harvard University and the Smithsonian Astrophysical Observatory.


\begin{thebibliography}{99}
\bibitem{note}
We use the following spectroscopic constants of CH$_2$ to parametrize Eq. (\ref{Hmol}) (in units of cm$^{-1}$): $A=73.06$, $B=8.42$, $C=7.22$, $D=0.778$, and $E=0.0399$ \cite{Bunker}.
\end{thebibliography}
\end{document}